\documentclass[11pt,a4paper]{article}
\usepackage{amsfonts}
\usepackage{amsmath}
\usepackage{amssymb}
\usepackage{jheppub}
\usepackage{ulem}
\usepackage{extarrows}
\usepackage{multirow}
\usepackage{float}

\usepackage{inputenc}
\usepackage{xspace}
\usepackage{url}

 \def\be{\begin{equation}}
\def\ee{\end{equation}}
\def\bea{\begin{eqnarray}}
\def\eea{\end{eqnarray}}

\def\a{\alpha}
\def\b{\beta}
\def\g{\gamma}

\def\r{\rho}
\def\t{\tau}

\def\s{\sigma}

 \def\r{\rightarrow}

 \def\IZ{\relax\ifmmode\mathchoice
 {\hbox{\cmss Z\kern-.4em Z}}{\hbox{\cmss Z\kern-.4em Z}}
 {\lower.9pt\hbox{\cmsss Z\kern-.4em Z}}
 {\lower1.2pt\hbox{\cmsss Z\kern-.4em Z}}\else{\cmss Z\kern-.4em Z}\fi}
 \def\IB{\relax{\rm I\kern-.18em B}}
 \def\IC{{\relax\hbox{$\inbar\kern-.3em{\rm C}$}}}
 \def\Ic{{\relax\hbox{$\inbar\kern-.22em{\rm c}$}}}
 \def\ID{\relax{\rm I\kern-.18em D}}
 \def\IE{\relax{\rm I\kern-.18em E}}
 \def\IF{\relax{\rm I\kern-.18em F}}
 \def\IG{\relax\hbox{$\inbar\kern-.3em{\rm G}$}}
 \def\IGa{\relax\hbox{${\rm I}\kern-.18em\Gamma$}}
 \def\IH{\relax{\rm I\kern-.18em H}}
 \def\II{\relax{\rm I\kern-.18em I}}
 \def\IK{\relax{\rm I\kern-.18em K}}
 \def\IP{\relax{\rm I\kern-.18em P}}

\def\Tr{{\rm Tr}}
 \font\cmss=cmss10 \font\cmsss=cmss10 at 7pt
 \def\IR{\relax{\rm I\kern-.18em R}}

\def\cH{{\cal H}}

\def\hL{{\hat{L}}}

\def\dd{\mbox{d}}

\def\a{\alpha}
\def\b{\beta}

\def\dd{\partial}
\def\D{\Delta}

\def\g{\gamma}

\def\k{\kappa}

\def\m{\mu}
\def\n{\nu}
\def\s{\sigma}

\def\r{\rho}
\def\t{\tau}

\def\O{{\cal O}}

\renewcommand{\@}[1]{\sqrt{#1}}
\renewcommand{\le}[1]{\label{#1}\end{eqnarray}}

\def\ffract#1#2{\raise .35 em\hbox{$\scriptstyle#1$}\kern-.25em/
\kern-.2em\lower .22 em \hbox{$\scriptstyle#2$}}

\def\hL{{\hat{L}}}

\newdimen\tableauside\tableauside=1.0ex
\newdimen\tableaurule\tableaurule=0.4pt
\newdimen\tableaustep
\def\phantomhrule#1{\hbox{\vbox to0pt{\hrule height\tableaurule width#1\vss}}}
\def\phantomvrule#1{\vbox{\hbox to0pt{\vrule width\tableaurule height#1\hss}}}
\def\sqr{\vbox{%
  \phantomhrule\tableaustep
  \hbox{\phantomvrule\tableaustep\kern\tableaustep\phantomvrule\tableaustep}%
  \hbox{\vbox{\phantomhrule\tableauside}\kern-\tableaurule}}}
\def\squares#1{\hbox{\count0=#1\noindent\loop\sqr
  \advance\count0 by-1 \ifnum\count0>0\repeat}}
\def\tableau#1{\vcenter{\offinterlineskip
  \tableaustep=\tableauside\advance\tableaustep by-\tableaurule
  \kern\normallineskip\hbox
    {\kern\normallineskip\vbox
      {\gettableau#1 0 }%
     \kern\normallineskip\kern\tableaurule}%
  \kern\normallineskip\kern\tableaurule}}
\def\gettableau#1 {\ifnum#1=0\let\next=\null\else
  \squares{#1}\let\next=\gettableau\fi\next}

\newcommand{\auth}{Institute of Theoretical Physics, Aristotle University of Thessaloniki, 54124 Thessaloniki, Greece}
\newcommand{\rminfn}{I.N.F.N. Sezione di Roma ``Tor Vergata", Via della Ricerca Scientifica, 00133 Roma, Italy}

\allowdisplaybreaks
\title{The free $\sigma$CFTs}
\author[a]{Andrea L. Guerrieri,}
\author[b]{\,Anastasios C. Petkou,}
\author[a]{and \,Congkao Wen}
\affiliation[a]{\rminfn}
\affiliation[b]{\auth}

\emailAdd{andrea.guerrieri@roma2.infn.it}
\emailAdd{petkou@physics.auth.gr}
\emailAdd{Congkao.Wen@roma2.infn.it}
\abstract{We introduce the conformal field theories that describe the shadows of the lowest dimension composites made out of massless free scalars and fermions in $d$ dimensions. We argue that these theories can be consistently defined as free CFTs for even $d\geq 4$. We use OPE techniques to study their spectrum and show that  for $d\rightarrow\infty$  it matches that of free bosonic CFTs in $d=6$ and $d=4$ dimensions. For these $\sigma$CFTs we calculate  $c_T$ in $d=6,8,10$ and $12$ dimensions using the OPE and also a direct construction of their higher-derivative energy momentum tensors.  Our results agree with the general proposal of arXiv:1601.07198. 
 }

\begin{document}
\maketitle
\section{Introduction}
Free conformal field theories (CFTs) can be  intriguing. This is one of the main lessons higher-spin holography\footnote{See for example the review \cite{Vasiliev:2012vf}  that also contains most of the relevant earlier references in higher-spin theories. For complementary approaches see e.g. \cite{Jin:2015aba} and references therein. } has taught us. For all their apparent simplicity, free CFTs emerge holographically as the result of a complicated symmetry that makes possible the consistent interaction of  infinite towers of higher-spin gauge fields. And even though symmetry breaking may be the most physically relevant property of a theory, the exploitation of its symmetric phase usually yields interesting knowledge.  

In this work we initiate the study of two peculiar free CFTs that are universally connected to free massless scalars $\phi(x)$ and Dirac fermions $\psi(x)$ in $d$ dimensions. For those free theories, the simplest  composite fields one can construct are\footnote{Spinor index contraction is tacitly assumed.} $\phi^2(x)$ and $\bar\psi(x)\psi(x)$ and parametrize the usual mass deformations. Considering then $x$-dependent mass terms such as $\int  d^dx \sigma_2(x)\phi^2(x)$ and $\int d^dx\sigma_1(x)\bar\psi(x)\psi(x)$ one elevates $\s_2(x)$  and $\s_1(x)$ to genuine scalar fields that determine the effective action of the massive phase. Moreover, since the canonical scaling dimensions of $\phi(x)$ and $\psi(x)$ are  $\D_\phi=\frac{d}{2}-1$ and $\D_\psi=\frac{d-1}{2}$, the fields $\sigma_2(x)$ and $\s_1(x)$ exhibit universal scaling behaviour in any $d$, namely $\D_{\s_2}=2$ and $\D_{\s_1}=1$. In the CFT language $\s_2(x)$ and $\s_1(x)$ are the {\it shadows}\footnote{See \cite{Metsaev:2008fs} for a nice discussion.} of the composites $\phi^2(x)$ and $\bar\psi(x)\psi(x)$ respectively. 

The fields $\s_2(x)$ and $\s_1(x)$ are not part of the spectrum of the corresponding  free CFTs for any $d$. They  show up, however, in the spectrum of  non-trivial bosonic and fermionic CFTs in $2<d<6$ \cite{Petkou:1994ad,Petkou:1996np,Fei:2014yja,Fei:2014xta} where they generically acquire anomalous dimensions. Remarkably, taking the strict $N\rightarrow\infty$ limit of the non-trivial theories one is left with theories where all fields have canonical dimensions and yet $\s_2(x)$ and $\s_1(x)$  are still in the spectrum replacing $\phi^2(x)$ and $\bar\psi(x)\psi(x)$ respectively.   Nevertheless, at least for the above range of dimensions, it has been noticed that all the non-trivial CFT models do become  free field theories for $d=2,4,6$ and hence it is natural to ask about the fate of the $\s$-fields in these cases. For the bosonic $O(N)$ vector model in $2<d<4$ this has been partially discussed in \cite{Leigh:2012mz}. It was there shown  that if one starts with both of the fields $\phi^2(x)$ and $\s_2(x)$ in the theory, the second one decouples in the free field theory limit and the first one becomes redundant at the non-trivial fixed point. This mechanism was connected to the  $O(N)\mapsto O(N-1)$ global symmetry breaking pattern. Moreover, $\s_2(x)$ decouples completely from the spectrum for $d=2$, while it coincides with $\phi^2(x)$ for $d=4$.   For the same model in $4<d<6$ one expects a similar\footnote{See \cite{Mati:2016wjn, Mati:2014xma} for some parallel developments.} picture, but now $\s_2(x)$ becomes indistinguishable from the elementary scalar $\phi(x)$ in $d=6$. The dimension of $\s_2(x)$ drops below the unitarity bound for $d>6$, which appears to imply that non-trivial bosonic CFTs in $d>6$ necessarily contain ghosts in the spectrum. In fermionic CFTs the relevant range of dimensions is $2<d<4$. For $d=2$, the $\s_1(x)$ field coincides with $\bar\psi(x)\psi(x)$, while for $d=4$ it looks very much like an elementary scalar  $\phi(x)$. For $d>4$ it falls below the unitarity bound.

Very recently a remarkable observation regarding the CFTs of the $\s$-fields - the $\s$CFTs - was made in\footnote{See also \cite{Stergiou:2015roa} for an earlier discussion.} \cite{Diab:2016spb,Giombi:2016fct}. In that work the parameter $C_T(d)$ that determine the two-point function of the energy momentum tensor \cite{Osborn:1993cr} in a generic CFT was studied in $N$-vector bosonic and fermionic models for general $d$. At the non-trivial fixed points of these models $C_T(d)$ can be calculated in a systematic $1/N$ expansion. For the bosonic $O(N)$ vector model the original calculation was done in \cite{Petkou:1994ad} using some form of the conformal bootstrap based on the OPE and consistency conditions. The result can generically be written as
\be
\label{CTb}
C_T^b(d)=C_T^{(0)}(d)\left[N+\frac{c_T(d)}{C_T^{(0)}(d)}+O(1/N)\right] \, ,
\ee
where $C_T^{(0)}(d)$ is the result corresponding to a single conformal free scalar field
\be
\label{CTb0}
C_T^{(0)}(d)=\frac{d}{(d-1)S_d^2}\,,\,\,\,S_d=\frac{2\pi^{d/2}}{\Gamma(d/2)} \, .
\ee
For the fermionic theory it was found in \cite{Diab:2016spb}
\be
\label{CTf}
C_T^f(d)={\cal C}_T^{(0)}(d)\left[N+\frac{\tilde{c}_T(d)}{{\cal C}_{T}^{(0)}(d)}+O(1/N) \right] \, ,
\ee
where ${\cal C}_T^{(0)}(d)$ is the result corresponding to a single conformal free Dirac fermion 
\be
\label{CTf0}
{\cal C}_T^{(0)}(d)=\frac{d}{2}\frac{1}{S_d^2}\Tr\mathbb{I} \, ,
\ee
in a spinor representation with dimension $\Tr\mathbb{I}$. The explicit formulae for $c_T(d)$ and $\tilde{c}_T(d)$ are rather complicated, but for even $d$ they considerably simplify to the following  expressions
\be
\label{ctcd}
c_T(d)=\frac{(-1)^{\frac{d}{2}+1}d(d-4)(d-2)!}{(d-1)\left(\frac{d}{2}+1\right)!\left(\frac{d}{2}-1\right)! S_d^2}\,,\,\,\,\,\,\tilde{c}_T(d)=\frac{(-1)^{\frac{d}{2}}d(d-2)(d-2)!}{2\left(\frac{d}{2}+1\right)!\left(\frac{d}{2}-1\right)! S_d^2} \, ,
\ee
and represent universal { corrections to $C_T(d)$ in the large-$N$ limit. 
The intriguing proposal then of \cite{Diab:2016spb} was that $c_T(d)$ and $\tilde{c}_T(d)$ are
the normalization coefficients of the energy momentum tensor two-point functions of two distinct  free $d$-dimensional CFTs. Namely,  $c_T(d)$ corresponds to the free CFT of a single $\sigma_2(x)$ field  and $\tilde{c}_T(d)$ to the free CFT  of a single $\sigma_1(x)$ field.

Motivated by the discussion above we initiate here the study of the free $\s$CFTs in general dimensions $d$. Our work is organized as follows. In the next section \ref{section:OPEnormal} we review in some detail the OPE analysis of normal free CFT in $d$ dimensions. This serves as the basis of our analysis of $\s$CFTs in Section \ref{section:sigmaCFT}, where we firstly argue that these theories can be consistently defined as free CFTs in even dimensions $d\geq 4$. We show that they have well-defined $d\rightarrow\infty$ limits in which they become indistinguishable from the usual free bosonic CFTs in $d=6$ (the $\s_2$CFT) or $d=4$ (the $\s_1$CFT). For even $d>6$ and $d>4$ the spectra of the $\s$CFTs involve, in principle, infinite towers of {\it ghost} and {\it null} higher spin fields with correspondingly negative and zero two-point function coefficients. Each tower is characretized by a generalized  twist\footnote{The standard definition of the twist $\t$ of an operator with dimension $\D_s$ and spin $s$ is $\t=\D_s-s$. For composite fields built from a free elementary conformal scalar $\phi(x)$ we generally have $\t=2(n+1)\D_\phi+2m$, $n,m=0,1,2, \ldots $. The integer $n$ defines the {\it trace-class} of the operator i.e. for $n=0$ we have single-trace operators, for $n=1$, double-trace etc. Within a trace-class, the even integer $2m$ defines the generalized twist $\tilde{\t}$. In this work twist will always stand for {\it generalized} twist.} $\tilde{\tau} =\D_s-s-2\D_\sigma$, where $\D_s$ is the dimension of the spin-$s$ fields in the tower and $\D_\s=\{\D_{\s_2},\D_{\s_1}\}$. In our case $\tilde\t=2k$, $k=0,1,2,..$. However, for {\it even} dimensions $d>6$ and $d>4$, only a finite number of towers contributes. The largest possible value of $\tilde\t$ is $\tilde{\t}_{max}=d-2-2\D_\s$, hence we have either $\tilde{\t}_{max}=d-6$ (for $\s_2$) or $\tilde{\t}_{\max}=d-4$ (for $\sigma_1$). The highest-twist towers contain now the energy momentum tensor and actually all higher-spin conserved currents of the free $\s$CFTs. The appearance of those higher-twist towers is crucial for the consistency of the conformal OPE with the free field theory calculation. We present a detailed OPE analysis of the four-point functions of the $\s$-fields and evaluate  $c_T(d)$ for $d=8,10$ and $\tilde{c}_T(d)$ for $d=6,8$.  Our results agree with those obtained from formulae (\ref{ctcd}).  In Section \ref{section:directCT} we present a direct approach to calculating $c_{T}(d)$ and $\tilde{c}_T(d)$ by a direct construction of the corresponding energy momentum tensors $T_{\m\n}(x)$. The latter now is a high-derivative composite field. Again our results coincide with those obtained from (\ref{ctcd}).\footnote{See also \cite{Osborn:2016bev}.} We conclude and discuss future directions in Section \ref{section:outlook}. Some technical details and a curious observation that relates $c_T(d)$ and $\tilde{c}_T(d)$ to the renormalization of composite operators are left to the Appendices.

\section{Normal free bosonic CFT's} \label{section:OPEnormal}
\subsection{The four-point function of the elementary scalar}

Since $\s_2(x)$ and $\s_1(x)$ are scalar fields, we would be interested in  free bosonic CFTs in $d$ dimensions. The unit normalized\footnote{In a CFT the two-point functions of all  quasi primary operators can be arbitrarily normalized, in which case the coefficients of three-point functions are the dynamically determined coupling constants.  This is not the case, however, for conserved currents where Ward identities relate two- and three-point functions.}  two-point function of the elementary field $\phi(x)$ in a bosonic {\it normal} free CFT$_d$ is given by 
\be
\label{2ptphi}
\langle\phi(x_1)\phi(x_2)\rangle = \frac{1}{x_{12}^{2\Delta_\phi}}\,.
\ee
Without  further input, the dimension $\D_\phi$ of $\phi(x)$ is at the moment arbitrary. All possible correlation functions of $\phi(x)$ and of all its composites, which constitute the full operator spectrum of the theory,  are calculated using standard Wick contractions. This trivial but non empty property constitutes the dynamics of free field theories.  
The field $\phi(x)$  also satisfies an {\it elementariness} condition that sets to zero its three-point function despite the fact that a non-zero result is allowed by conformal invariance. This appears to be part of generalized partity-even condition under $\phi(x)\mapsto -\phi(x)$ satisfied by all correlation functions of the free theory.  Its four-point function is given by
\be
\label{4ptphi}
\langle \phi(x_1)\phi(x_2)\phi(x_3)\phi(x_4)\rangle = \frac{1}{(x_{12}^2x_{34}^2)^{\Delta_\phi}}\left[1+v^{\Delta_\phi}\left(1+\frac{1}{(1-Y)^{\Delta_\phi}}\right)\right] \, ,
\ee
and it depends on two conformally invariant ratios. Here we are using $v$ and $Y$ that are relevant for the short distance limits $x_{12}^2,x_{34}^2\rightarrow 0$, and therefore for the direct channel OPE
\be
\label{confratios}
u=\frac{x_{12}^2x_{34}^2}{x_{13}^2x_{24}^2}\,,\,\,\,v=\frac{x_{12}^2x_{34}^2}{x_{14}^2x_{23}^2}\,,\,\,\,Y=1-\frac{v}{u} \, .
\ee
The main CFT condition on the normal free theory is that all its correlation functions  can be expanded and matched with the conformal OPE of the fields involved. Precisely, the four-point function (\ref{4ptphi}) can be expanded as 
\be
\label{OPE4ptphi}
\Phi(x_1,x_2,x_3,x_4) = \frac{1}{(x_{12}^2x_{34}^2)^{\Delta_\phi}}\left[1+\frac{g_{\O_2}^2}{C_{\O_2}}{\cal H}_{\Delta_{\O_2}}(v,Y) +\sum_{k=1}^{\infty}\frac{g_{T_{2,2k}}^2}{C_{T_{2,2k}}}{\cal H}_{\Delta_{T_{2,2k}}}(v,Y)+\cdots\right] \, .
\ee
Here ${\cal H}_{\D_{\O_2}}(v,Y)$ is the conformal partial wave (CPW) of a scalar operator $\O_2$ with dimension $\D_{\O_2}=2\D_\phi$ and ${\cal H}_{T_{2,2k}}(v,Y)$ are the CPWs of symmetric and traceless tensors $T_{2,2k}$ of rank-$2k$ with $k=1,2, \ldots $. The latter correspond to operators with spin $s=2k$, $k=1,2, \ldots $ and dimensions $\Delta_{T_{2,2k}}=\D_{\O_2}+2k$. $C_{T_{2,2k}}$ are the normalization constants of the two-point functions of $T_{2,2k}$ and  $g_{T_{2,2k}}$ are the coefficients of their three-point functions with two ${\O}_2$'s.  As far as the OPE analysis is concerned only the ratios $g^2_{T_{2,2k}}/C_{T_{2,2k}}$ are relevant. However, we need to be careful when conserved currents satisfying Ward identities are involved, as for example the energy momentum tensor $T_{\m\n}$ which is the unique
spin-2, and dimension $\D_T=d$ operator that  appears in the OPE. We recall in Appendix~\ref{AppA} some  useful formulae  for the scalar  $\O_2$ and  $T_{\m\n}$.  

\subsection{The CPWs}
The fields depicted on the rhs of (\ref{OPE4ptphi}) constitute the {\it leading} generalized twist higher spin tower. They all have $\tilde\t =0$ and as we will see below they all contribute a power of $v^{\D_\phi}$ in the four-point function. Higher-twist fields would  necessarily contribute higher powers of $v$, and their apparent absence in the free field result (\ref{4ptphi}) needs to be explained. This is what we do below.

 The generic CPW for a scalar field with dimension $\D$  is known for all $d$. It can be written as a double series in $v$ and $Y$ as
\begin{align}
\label{Hscalar}
{\cal H}_{\D}(v,Y;d)&=v^{\frac{\D}{2}}\left[F_\D^{0}(Y)+\sum_{n=1}^{\infty}\frac{v^n}{n!}\frac{\left(\frac{1}{2}\D\right)_n^4}{(\D)_{2n}(\D+1-\frac{d}{2})_n}F_\D^n(Y)\right] \, ,
\end{align}
with $(a)_n$ the Pochhammer symbol. We will be using henceforth a compact notation for the particular form of the hypergeometric functions that  generically appears in spin-$s$ CPWs 
\be
\label{2F1}
F_a^k(Y)\equiv {}_2F_1\left(\frac{a}{2}+k,\frac{a}{2}+k;a+2k;Y\right)\,,
\,\,\, F_{a+2}^k(Y)=F_a^{k+1}(Y)\,\,\, {\rm with} \,\,\, k=0,1,2,3, \ldots \, .
\ee 
Notice now that only the second term in (\ref{Hscalar}) carries an explicit $d$ dependence, which appears in the denominators and it is correlated with the powers of $v$. This implies that (\ref{Hscalar}) may be cast in the suggestive form
\be
\label{Hscalar1/d}
{\cal H}_{\D}(v,Y;d)=v^{\frac{\D}{2}}\left[F_\D^{0}(Y)+\sum_{n=1}^{\infty}\left(\frac{v}{d}\right)^nH_{\D}^{(n)}(Y,d)\right] \, ,
\ee
with $H_\D^{(n)}(Y,d)$ some complicated but known functions that are analytic as $Y\rightarrow 0$ and $d\rightarrow\infty$. There are two issues with the form (\ref{Hscalar1/d}) of the scalar CPW. Firstly, for unitary CFTs it holds $\D\geq \frac{d}{2}-1$ and hence the expansion (\ref{Hscalar1/d}) is not very useful.\footnote{The large-$d$ behaviour of  CPWs for unitary CFTs was studied in\cite{Fitzpatrick:2013sya}.} It could be useful however for fixed $\D$, and hence for the $\s$CFTs. But this brings in a second issue since the factors $\left(\D_\phi+1-\frac{d}{2}\right)_n$ could develop poles as one  scans through the dimension $d$.  If the latter is resolved, and we will see how this can happen in the next Section, then (\ref{Hscalar1/d}) demonstrates that for {\it fixed} $\D$ the natural expansion parameter for the scalar CPWs is $v/d$ and this has some interesting implications.  For example, ${\cal H}_{\D}(v,Y;d)$ would then simplify considerably in the strict $d=\infty$ limit giving the  finite and $d$ independent result $F_\D^0(Y)$.

The CPWs for arbitrary even spin-$s\geq 2$ tensors and for general $d$ were known for a long time \cite{Hoffmann:2000mx}. They are considerably more complicated but and important tiding up was achieved by Dolan and Osborn, starting with \cite{Dolan:2000ut,Dolan:2003hv} and continued with subsequent work by several other groups (see for example \cite{Costa:2011dw}).  Some of the formulae that we will be using here are generalizations of those very useful works.   Results for general $d$ are not very practical unless one identifies certain patterns in them. For a given spin-$s$ field the pattern follows the finite expansion of the Genenbauer polynomial $C_{s}^{d/2-1}(t)$ as it is explained in the Appendix~\ref{AppB}. Our approach is exemplified by the spin-2, 4 and 6 cases, which suffice for the calculations in Section~\ref{section:sigmaCFT}.  The CPW for spin-2, 4 and 6 fields with dimensions $\D_2,\D_4$  and $\D_6$ take the form
\begin{align}
\label{H2}
\cH_{\D_2}(v,Y;d) &=v^{\frac{\D_2}{2}-1}\sum_{k=0}^\infty v^k\kappa_{\D_2,2k}\left[Y^2F_{\D_2+2}^{k}(Y)-4\frac{v}{d}\alpha^{(1)}_{\D_2,2k}F_{\D_2}^{k}(Y)\right]\, ,\\
\label{H4}
\cH_{\D_4}(v,Y;d)&=v^{\frac{\D_4}{2}-2}\sum_{k=0}^\infty v^{k}\kappa_{\D_4,2k}\left[Y^4F_{\D_4+4}^{k}(Y)-4\frac{6vY^2}{d+4}\alpha^{(1)}_{\D_4,2k}F_{\D_4+2}^{k}(Y)\right.\cr
&\hspace{4.5cm}\left.+4^2\frac{3v^2}{(d+2)(d+4)}\alpha^{(2)}_{\D_4,2k}F_{\D_4}^{k}(Y)\right]\,,
\\
\label{H6}
\cH_{\D_6}(v,Y;d)&=v^{\frac{\D_6}{2}-3}\sum_{k=0}^\infty v^{k}\kappa_{\D_6,2k}\left[Y^6F_{\D_6+6}^{k}(Y)-4\frac{15vY^2}{d+8}\alpha^{(1)}_{\D_6,2k}F_{\D_6+4}^{k}(Y)\right.\cr
&\hspace{.7cm}+\left.4^2\frac{45v^2Y^2}{(d+8)(d+6)}\alpha^{(2)}_{\D_6,2k}F_{\D_6+2}^{k}(Y)-4^3\frac{15v^3}{(d+4)(d+6)(d+8)}\alpha^{(3)}_{\D_6,2k}F_{\D_6}^k(Y)\right].
\end{align}
The coefficients $\kappa_{\D_n,m}$ and $\alpha_{\D_n,m}^{(i)}$ for $n,m=2,4,6, \ldots $, $i=1,2,3, \ldots $ are functions of $\D_2,\D_4, \D_6$ and $d$. We also note that $\kappa_{\D_n,0}=\alpha_{\D_n,0}^{(i)}=1$. Since the aim of our present work is to introduce the $\s$CFTs, we did not attempt to find their form in complete generality but only for $s=2,4$ that are needed in Section~\ref{section:sigmaCFT}. These were obtained relatively easily using Mathematica and are presented in the Appendix~\ref{AppC} and~\ref{AppD}, indicating that a dedicated but straightforward numerical work should yield the general formulae.\footnote{After our work appeared, we have learned from the revised version of \cite{Osborn:2016bev} that analytic expression for the OPE of free four-point functions had previously appeared in \cite{Fitzpatrick:2011dm}. It would be very interesting to make a direct comparison with the latter reference.} Our results imply further that the CPW for spin-$s$ fields can also be written in a form analogous to (\ref{Hscalar1/d}), namely
\be
\label{Hspin1/d}
{\cal H}_{\D_s}(v,Y;d)=v^{\frac{\D_s-s}{2}}\left[Y^s F_{\D_s+s}^{0}(Y)+\sum_{n=0}^\infty\left(\frac{v}{d}\right)^nH_{\D_s}^{(n)}(Y,d;s)\right] \, ,
\ee
where the complicated functions $H_{\D_s}^{(n)}(Y,d;s)\rightarrow Y^s$ as $Y\rightarrow 0$ and are analytic as $d\rightarrow\infty $. We see again that  $v/d$ emerges as the relevant expansion parameter and for fixed $\D_s$ only the first term on the RHS of (\ref{Hspin1/d}) survives in the $d\rightarrow\infty$ limit.

\subsection{OPE analysis of normal free CFTs} \label{section:OPEfree}

The above general results  are relevant for the OPE analysis of the four-point function (\ref{4ptphi}).
We have  claimed that only the leading-twist higher-spin fields appear in (\ref{OPE4ptphi}) and this is confirmed by virtue of the validity of the following expansion
\be
\label{sumd0}
1+\frac{1}{(1-Y)^{\D_\phi}}=\sum_{s=0,2,4, \ldots }^{\infty}\a_s(\D_\phi)Y^{s}F_{2\D_\phi+2s}^{0}(Y) \, ,
\ee
with
\be
\label{alphas}
\alpha_s(\Delta_\phi)\equiv \frac{g_{T_{2,s}}^2}{C_{T_{2,s}}}=\frac{(\Delta_\phi)_{s/2}(\Delta_\phi)_s}{2^{s-1}s!\left(\Delta_\phi+\frac{s-1}{2}\right)_{s/2}}\,.
\ee
Namely, the free four-point function is constructed exclusively from the leading $d$ independent part of the higher-spin current CPWs. As a result, only $v^{\D_\phi}$ appears in (\ref{4ptphi}) which asks the question how all other powers of $v$ have disappeared. To uncover the mechanism, since  $\D_{\O_2}=2\D_\phi$ we see  from (\ref{Hspin1/d}) that the $v^{\D_\phi +1}$ term can only appear in the OPE as the result of the scalar and the spin-2 CPW contributions. From (\ref{Hscalar}), (\ref{H2})
and the expressions (\ref{k2}), (\ref{alpha2}) in the Appendix~\ref{AppC}
this is found to be
\be
\label{vDphi1}
v^{\D_\phi+1}\left[\frac{g_2^2}{C_{2}}\frac{\D_\phi^3}{8(\D_\phi+1)(2\D_\phi+2-d)}F_{2\D_\phi}^1(Y)-\frac{g_{T_{2,2}}^2}{C_{T_{2,2}}}\frac{4}{d}F_{2\D_\phi+2}^0(Y)\right] \, .
\ee
To proceed we need from (\ref{alphas}) the following expressions		
\be
\label{free_results}
\frac{g_2^2}{C_{2}}=2\,,\,\,\,\frac{g_{T_{2,2}}^2}{C_{T_{2,2}}}=\frac{\D_\phi^2(\D_\phi+1)}{2(2\D_\phi+1)} \, .
\ee
Then, using  the formulae in the Appendix \ref{AppC} we find from (\ref{Hscalar}) and (\ref{H2}) that the OPE should contain the following term
\be
\label{cc}
v^{\D_\phi+1}\frac{2\D_\phi^2(d-2-2\D_\phi)}{d(4\D_\phi+2-d)}F_{2\D_\phi}^1(Y) \, .
\ee
This term should vanish and for that we see three possibilities that  define  respectively three types of free CFTs. We discuss in this Section one of them, leaving the other two for Section \ref{section:sigmaCFT}. The simplest possibility  is to consider the vanishing of (\ref{cc}) as a condition on $\D_\phi$. This yields $\Delta_\phi=\frac{d}{2}-1$, which is the usual canonical value for the dimension of the elementary unitary conformal free scalar $\phi$ field in any dimension.  Now everything falls in place, and the field $T_{2,2}$ is identified with the energy momentum tensor $T_{\m\n}$ having canonical dimension $\D_{T_{2,2}}\equiv\D_T=d$, while the usual Ward identity and free-field theory results for $C_T$ follow
\be
\label{gphiphiT}
\frac{g_{T_{2,2}}^2}{C_{T_{2,2}}}\equiv \frac{1}{4}\frac{g_T^2}{C_T}\,,\,\,\,\,g_{T}=\frac{d\Delta_\phi}{(d-1)S_d}\,,\,\,\,C_T=\frac{d}{(d-1)S_{d}^2}\,.
\ee
The scalar field in the OPE (\ref{OPE4ptphi}) can be identified with the (normalized) composite field 
\be
\label{O}
{\cal O}_2(x)=\frac{1}{\sqrt{2}}\phi(x)\phi(x) \, .
\ee
Terms proportional to $v^{\D_\phi+k}$ come from the CPW contributions of the $k$ fields with spins $s=0,2,4, \ldots , 2k.$, and they all vanish\footnote{We are not aware of general proof for this statement, however it it easily checked on a computer, to high enough orders, using the explicit expressions for the coefficients of the CPWs.}  when $\D_\phi$ is set to its free canonical value above. In this case too  the dimensions of the higher-spin fields become $\D_s=d-2+s$ and correspond to the usual leading-twist higher-spin conserved currents.

A similar OPE analysis can be done for four-point functions of composite fields. The simplest example is that of the scalar field $\O_2(x)$ the has been found in the four-point function of $\phi(x)$. 
We have
\begin{align}
\label{4ptO}
\langle \O_2(x_1) \ldots \O_2(x_4)\rangle &= \frac{1}{(x_{12}^2x_{34}^2)^{2\Delta\phi}}\left[1+v^{2\Delta_\phi}\left(1+\frac{1}{(1-Y)^{2\Delta_\phi}}+\frac{4}{(1-Y)^{\Delta_\phi}}\right)+ 4v^{\Delta_\phi}\left(1+\frac{1}{(1-Y)^{\Delta_\phi}}\right)\right]\, .
\end{align}
The last term in (\ref{4ptO}) gives  by virtue of  (\ref{sumd0}) the contribution of the leading-twist higher-spin conserved currents. We also find
\be
\label{sumd02tr}
\sum_{s=0,2,4, \ldots }^{\infty}\left[\a_s(2\D_\phi)+\beta_s(2\D_\phi)\right]Y^{s}F_{4\D_\phi+s}^0(Y)=1+\frac{1}{(1-Y)^{2\D_\phi}}+\frac{4}{(1-Y)^{\D_\phi}}\, ,
\ee
where $\alpha_s(2\D_\phi)$ is given by (\ref{alphas}) and $\beta_s(2\D_\phi)$ by
\be
\label{betas}
\beta_s(2\D_\phi) =\frac{\left(\D_\phi\right)_{s/2}^2}{2^{s-2}\left(s/2\right)!\left(2\D_\phi+\frac{s-1}{2}\right)_{s/2}}\,,\,\,\,s=0,2,4, \ldots \, .
\ee
By virtue of the discussion in the previous subsection, we see that (\ref{sumd02tr}) gives the contribution of  fields $T_{4,2k}$ with even spin $s=2k$, $k=0,1,2, \ldots $ and dimensions $\D_{4,2k}=4\D_\phi+2k$. These are  leading-twist  {\it double-trace} higher-spin fields. The corresponding ratios of  their three- to two-point functions then read
\be
\label{gC4}
\frac{g_{4,2k}^2}{C_{4,2k}}= \a_s(2\D_\phi)+\beta_s(2\D_\phi) \, .
\ee
As before, one can verify that the absence of higher order powers in $v$ results from the condition $\D_\phi=\frac{d}{2}-1$.  Analogous results can be found for the four-point function of all scalar operators of the form $\O_k(x)=c_k\phi^k(x)$, $k=3,4, \ldots$. One finds that the spectrum is composed out of leading-twist multi-trace operators.

\subsection{Beyond free unitary CFTs}

When $\D_\phi\neq \frac{d}{2}-1$ we need a different mechanism to deal with higher powers of $v$, such as $(\ref{cc})$. Firstly, in unitary theories  the {\it anomalous dimension} $\g_\phi=\D_\phi-\frac{d}{2}+1>0$ and (\ref{cc}) is non-zero and negative. Then, provided that we wish to stick to the free theory when the four-point function is still given by Wick contraction, the only way to cancel this term is by enhancing the OPE of $\phi(x)$. Namely, a term like (\ref{cc}) can be cancelled by the CPW contribution of a scalar field $\O_4(x)$ with dimension $\D_4=2\D_\phi+2$. This is a twist-two operator and its explicit form, fixed up to an arbitrary overall normalization,  is 
\be
\label{O4}
\O_4(x)=\frac{1}{2}\partial_\m\phi(x)\partial_\m\phi(x)-\frac{\D_\phi}{2+2\D_\phi-d}\phi(x)\partial^2\phi(x) \, .
\ee
One then finds
\be
\label{O423pt}
\langle\O_4(x_1)\O_4(x_2)\rangle =C_4\frac{1}{(x_{12}^2)^{2\D_\phi+2}}\,,\,\,\,\,\langle\phi(x_1)\phi(x_2)\O_4(x_3)\rangle = g_4\frac{x_{12}^2}{(x_{13}^2x_{23}^2)^{\D_\phi+1}} \, ,
\ee
where
\be
\label{C4g4}
C_4=\frac{2d\D_\phi^2(2+4\D_\phi-d)}{(2+2\D_\phi-d)}\,,\,\,\,g_4=2\D^2_\phi \, .
\ee
The CPW of the scalar $\O_4(x)$ CPW cancels (\ref{cc}), but now the bag of Aeolus has opened and in principle all towers of higher-spin $s=2,4,6, \ldots $ and higher-twist $\tilde\t=2,4,6, \ldots$ fields with dimensions $\D_s=2\D_\phi+\tilde\t+s$, need to be included in the $\phi(x)$ OPE in order to reproduce the free field theory result. In general,  in such a case none of the operators in the OPE can be identified with conserved higher-spin currents in $d$-dimensions, and in particular with a $d$-dimensional energy momentum tensor, unless there is a maximum value $\tilde\t_{\max}$ for which it holds
\be
\label{HSbreak}
2\D_\phi=d-2-\tilde\t_{\max} .
\ee
Namely, the dimension of $\phi(x)$ must fall below the unitarity bound. In this case, we need to add only a finite number of higher-twist, higher-spin towers, as the towers with  $\tilde\t>\tilde\t_{\max}$ enter the OPE with vanishing coefficients.  The tower with maximum twist $\tilde\t_{\max}$ contains the energy momentum tensor and all higher-spin conserved currents. Contrast this with the fact that in unitary CFTs the energy momentum tensor belongs to the leading-twist towers of higher-spin fields. 

If one wants to keep unitarity, the well known way out is of course to move beyond free field theory. The remarkable recent progress in the new bootstrap (see the recent review \cite{Rychkov:2016iqz} and references therein) is one way to proceed. A less explored, yet analytic, method \cite{Petkou:1994ad,Petkou:1996np} is to deform the free field theory results such as (\ref{4ptphi}) or (\ref{4ptO}) by conformal integrals in a way consistent with crossing symmetry.  These additional {\it skeleton} graphs introduce terms that can account for the presence of the infinite towers of higher-twist and higher-spin  operators in the spectrum. Matching  these terms with the OPE then fixes the anomalous dimensions and the coupling constants of the theory. This method is useful in the context of the $1/N$ expansion and we believe that it may give interesting new results when combined with the modern language of Mellin amplitudes \cite{Mack:2009mi, Fitzpatrick:2011ia}.

\section{The $\s$CFTs} \label{section:sigmaCFT}

\subsection{The free $\s$CFTs for finite $d$}

As we have seen above, if we insist on reproducing a free CFT keeping $d$ finite but taking $\D_\phi\neq\frac{d}{2}-1$ we are forced to consider non-unitary CFTs. This is exactly the case of the $\s$CFTs considered here. However it is not immediately clear whether one can consistently define $\s_2(x)$ and $\s_1(x)$ as free fields in any $d$. A free field needs to obey the elementariness condition that sets to zero its three-point function. In this context recall that the coupling $g_{\s_2}(d)$ of the $\s_2(x)$ three-point function was calculated for general $d$ in \cite{Petkou:1994ad} to be\footnote{We have normalized all two-point functions to one and set $N=1$.}
\be
\label{element}
g_{\s_2}(d)=2(d-3)g_*(d)\,,\,\,\,\,\,\,\,g_{*}^2(d)=\frac{2\Gamma(d-2)}{\Gamma\left(3-\frac{d}{2}\right)\Gamma^3\left(\frac{d}{2}-1\right)}
\ee
Remarkably, the poles in the denominator of the second formula in (\ref{element}) set  $g_{\s_2}(d)=0 $  for even dimensions $d\geq 6$. Similarly\footnote{The precise calculation has not been written down though.} $g_{\s_1}$ which is the three-point coupling of $\s_1(x)$ is proportional to the critical coupling $G_*(d)$ given in eq.$(37)$ of \cite{Petkou:1996np} and vanishes for even $d\geq 4$. This observation shows that at least for even $d\geq 4$ we can define the free $\s$CFTs in a way completely analogous to that of the normal free CFTs e.g. to evaluate correlation functions of $\s_2(x)$ and $\s_1(x)$ and their composites Wick contraction without the need of an underlying Lagrangian formalism.  The only difference with the normal free CFTs is the fact that the $\s$CFTs  are not unitary. One way in which the non-unitary character of the $\s$CFTs shows up is the fact that we encounter {\it ghost fields} (i.e. conformal fields with negative two-point function coefficients) in their spectrum. These field are necessary in order to cancel the higher powers of $v$ and reproduce the free field result. For example, the two-point function coefficients of the field $\O_2(x)$ that appears in the four-point functions of $\s_2(x)$ and $\s_1(x)$ read
\be
\label{O2s2s1}
C_4\Bigl|_{\s_2}=\frac{8d(10-d)}{6-d}\,,\,\,\,C_4\Bigl|_{\s_1}=\frac{2d(6-d)}{4-d} \, ,
\ee
and they are negative in the dimension ranges $6<d<10$ for $\s_2$CFT and $4<d<6$ for $\s_1$CFT. When $d>10$ and $d>6$, the field $\O_2(x)$ is not a ghost any more (although it is non-unitary), but higher-twist scalars take its place in the ghost spectrum. 

Another issue that has already been mentioned in Section~\ref{section:OPEnormal} is that for even dimensions we will encounter poles  in the CPWs such as (\ref{Hscalar}). However one has to consider the full OPE in order to correctly identify them. For example, as the discussion of the $\O_2(x)$ field in Section~\ref{section:OPEfree} shows, the relevant terms in the four-point functions of $\s_2(x)$ and $\s_1(x)$ are respectively
\be
\label{poles}
v^3\frac{8(d-6)}{d(10-d)}F_4^1(Y)\,,\,\,\,\,v^2\frac{2(d-4)}{d(6-d)}F_2^1(Y) \, .
\ee
We see then that the poles at $d=10$ and $d=6$ are exactly cancelled, because for these dimensions  $\O_2(x)$  becomes {\it null} i.e. its two-point function vanishes.  Notice that nevertheless the three-point function of $\O_2(x)$ with the $\s$ fields remains non zero. Hence, we observe an intriguing structure in the spectrum of the free $\s$CFTs in even dimensions, where ghost, null and normal conformal fields combine in a precise way to produce the free field theory result.
\subsection{The $d\rightarrow\infty$ limit}

Another  way to reproduce the free field theory result (\ref{4ptphi}) using the conformal OPE is to take the strict $d\rightarrow\infty$ limit. As we have seen, for fixed $\D_\phi$  this yields a non-trivial and $d$ independent result. Moreover, the contributions of leading-twist higher-spin fields with dimensions $\D_s=2\D_\phi+s$ suffice to reproduce (\ref{4ptphi}) and all other CPWs drop out. This observation holds for arbitrary $\D_\phi$ it implies that we can formally identify the spectrum of a normal free CFT in $d$-dimensions with the $d\rightarrow\infty$ limit of a non-unitary CFT whose elementary field has fixed dimension. This seems to be particularly fitting for the discussion of the $\s$CFTs of interest to us here. The fields $\s_2(x)$ and $\s_1(x)$ have fixed dimensions $\D_{\s_2}=2$ and $\D_{\s_1}=1$. From the one hand they can be identified with normal free scalars in dimensions $d=6$ and $d=4$ respectively.  In these dimensions the spectrum of the theories consists only of leading-twist higher-spin fields, which turn out to be also conserved currents. The spectrum of the $\s$CFTs changes completely for any other dimension $d$, but in the $d\rightarrow\infty$ limit it appears to settle down again back to leading-twist higher-spin fields. However, in the latter case we cannot talk about conserved currents since even the very notion of a field living in an infinite dimensional space is not well defined. Perhaps one can try to make sense of this observation by considering the $v/d$ corrections in a more systematic way. 
\subsection{$c_T$ for the $\s$CFTs}

We went at lengths above to argue that free $\s$CFTs for even $d$ can be described using CFT methods while keeping in their spectrum the infinite tower of higher-spin conserved currents which is the crucial characteristic of a free field theory.  In particular, these theories always possess an energy momentum tensor.  As a final example in our introduction of the free $\s$CFTs we will present here the explicit results for the coefficients $c_T$ in the two-point function of the energy momentum tensor of these theories, in dimensions $d=6,8,10$. Our results agree with those of \cite{Diab:2016spb, Osborn:2016bev}. 

For $\s_2(x)$ and $\s_1(x)$ the first non-trivial dimensions are $d=8$ and $d=6$. In these cases the field $\O_4(x)$ has dimensions $\D_4=6$ and $\D_4=4$ respectively. Therefore, the spin-2 fields in the same-twist towers have dimensions $\D_{4,2}=8$ and $\D_{4,2}=6$ and therefore they coincide with the energy momentum tensors in dimensions $d=8$ and $d=6$. Consider first the $\s_2(x)$ case where $\D_2=2$. The leading contribution of the energy momentum tensor in $d=8$ should be a term of the form $v^3Y^2$. 
This term is there to cancel the corresponding contributions that come from the leading-twist spin-2 and spin-4 CPWs, namely
\be
\label{d8s2calc}
\frac{g_{2,2}^2}{C_{2,2}}\kappa_{6,2}^{(1)}-\frac{g_{2,4}^2}{C_{2,4}}\frac{24}{d+4}+\frac{1}{4}\frac{g_T^2}{c_T}=0 \, ,
\ee
for $\D_2=4$ and $d=8$. Using the explicit expressions in the Appendix \ref{AppC} we obtain\footnote{For simplicity, here and in the following discussions we will drop the universal factor ${1 \over S_d^2}$ on $c_T$ and $\tilde{c}_T$.}
\be
\label{CTd8s2}
\frac{1}{4}\frac{g_T^2}{c_T}=-\frac{2}{7}\,\,\Rightarrow\,\,c_T=-\frac{32}{7} \, ,
\ee
which agrees with the first of (\ref{ctcd}). For $\s_1(x)$ with $\D_2=1$ and $d=6$ the corresponding formulae are 
\be
\label{d6s1calc}
\frac{g_{1,2}^2}{C_{1,2}}\kappa_{4,2}^{(1)}-\frac{g_{1,4}^2}{C_{1,4}}\frac{24}{d+4}+\frac{1}{4}\frac{g_T^2}{\tilde{c}_T}=0 \, ,
\ee
\be
\label{CTd6s1}
\frac{1}{4}\frac{g_T^2}{\tilde{c}_T}=-\frac{3}{50}\,\,\Rightarrow\,\,\tilde{c}_T=-6 \, ,
\ee
which agrees with the second of (\ref{ctcd}).
Going up to $d=10$ and $d=8$ increases  the complexity since now the energy momentum tensor is a twist-four field. Its leading behaviour for $\s_2(x)$ is $v^4Y^2$ and it is there to cancel corresponding contributions from spin-$s=2,4,6$ leading-twist fields and spin-$s=2,4$ twist-two fields. The explicit expression reads
\be
\label{d10s2calc}
\frac{g_{2,2}^2}{C_{2,2}}\kappa_{6,2}^{(2)}-\frac{g_{2,4}^2}{C_{2,4}}\kappa_{8,4}^{(1)}\alpha_{8,4}^1\frac{24}{d+4}+\frac{g_{2,6}^2}{C_{2,6}}\frac{720}{(d+6)(d+8)}+\frac{g_{4,2}^2}{C_{4,2}}\kappa_{8,2}^{(1)}-\frac{g_{4,4}^2}{C_{4,4}}\frac{24}{d+4}+\frac{1}{4}\frac{g_T^2}{c_T}=0 \, , 
\ee
which we need to evaluate for $d=10$. We obtain
\be
\label{CTd10s2}
\frac{1}{4}\frac{g_T^2}{c_T}=\frac{5}{63}\,\,\Rightarrow\,\,c_T=\frac{140}{9} \, ,
\ee
which agrees with the first of (\ref{ctcd}). For  $\s_1(x)$ we calculate the analogous expression
\be
\label{d8s1calc}
\frac{g_{1,2}^2}{C_{1,2}}\kappa_{4,2}^{(2)}-\frac{g_{1,4}^2}{C_{1,4}}\kappa_{6,4}^{(1)}\alpha_{6,4}^1\frac{24}{d+4}+\frac{g_{2,6}^2}{C_{2,6}}\frac{720}{(d+6)(d+8)}+\frac{g_{4,2}^2}{C_{4,2}}\kappa_{6,2}^{(1)}-\frac{g_{4,4}^2}{C_{4,4}}\frac{24}{d+4}+\frac{1}{4}\frac{g_T^2}{\tilde{c}_T}=0 \, ,
\ee
for $d=8$. We obtain
\be
\label{CTd8s1}
\frac{1}{4}\frac{g_T^2}{\tilde{c}_T}=\frac{2}{147}\,\,\Rightarrow\,\,\tilde{c}_T=24 \,, 
\ee
which agrees with the second of (\ref{ctcd}). In the following section, we will compute two-point functions of higher-derivative energy momentum tensors, from which we will confirm the results obtained in this section by CPW, and even to obtain new results for $c_T$ of $\s_2$ in $d=12$ as well as $\tilde{c}_T$ of $\s_1$ in $d=10$. 

\section{The direct calculation for $c_T$ in $\s$CFTs} \label{section:directCT}
In this section, we will perform a direct computation on $c_T$ by computing the two-point functions of energy momentum tensors $T_{\mu \nu}$. This direct method allows us to obtain $c_T$ quite straightforwardly for energy momentum tensors up to eight derivatives in any dimension. Specify to scalars with conformal dimensions $\Delta=2$ and $\Delta=1$, the results lead to $c_T$ for $d=8, 10, 12$ and $\tilde{c}_T$ for $d=6, 8, 10$, respectively. 
For that we need explicit expressions for the free energy momentum tensors. This can actually be done without any use of Lagrangian as their general form can be fixed by symmetry, traceless, conservation as well as matching with the Ward identity result (\ref{gT}). 
For convenience, we mark the energy momentum tensors by the number of derivatives acting on the elementary field $\s(x)$. The simplest case is that of the canonical two-derivative energy momentum tensor of a normal free scalar theory, 
\bea 
T^{(2)}_{\mu \nu} (x)=\partial_{\mu} \sigma(x) \partial_{\nu} \sigma(x) - {d-2 \over 4(d-1)} \partial_{\mu} \partial_{\nu} \sigma^2(x)
-{1 \over 2(d-1)} \eta_{\mu \nu} \partial \sigma(x) \cdot \partial \sigma(x) \, ,
\eea
where the superscript $(2)$ indicates two derivatives. The scalar $\sigma(x)$ has  canonical dimension $\Delta_{\sigma} = {d \over 2} -1$ such that $T^{(2)}_{\mu \nu}$ has dimension $d$ as required. Although the above energy momentum tensor may be derived from the usual free scalar Lagrangian, 
it can also be found by  writing down all possible symmetric tensors with two derivatives and two scalar fields $\sigma$, and trying to fix all the  coefficients by tracelessness, conservation and imposing the Ward identity.  This is done on-shell using the equations of motion for the free massless field $\s(x)$. As before, when the dimension of $\s(x)$ is canonical the equation of motion is always quadratic, namely $\partial^2 \sigma(x)=0$. But for fixed dimension $\D$ the equation of motion is generically higher derivative, with the number of derivatives depending on $d$. Specifically, the massless free equation scalar of motion is found from
\be
\label{eom}
\partial^{2n}\langle\sigma(x)\s(0)\rangle=\partial^{2n}\frac{N_d}{x^{2\D}}=4^n(\D)_n(\D+1-\frac{d}{2})_n\frac{N_d}{x^{2(\D+n)}}=0 \, .
\ee
where the normalization constant $N_d$ is defined below. Therefore, the equation of motion $\partial^{2n}\s(x)=0$ is relevant for a scalar field with scaling dimension $\D=\frac{d}{2}-n$. 

In the two-derivative  case, up to the equation of motion there are three possible symmetric tensors thus we can generally write
\bea 
T^{(2)}_{\mu \nu} (x)=c_1 \partial_{\mu} \sigma(x) \partial_{\nu} \sigma(x) + c_2 \partial_{\mu} \partial_{\nu} \sigma^2(x)
+ c_3 \eta_{\mu \nu} \partial \sigma(x) \cdot \partial \sigma(x) \, .
\eea
The traceless and conservation conditions lead to two constraints on the parameters, which are given as 
\bea
c_1 + 2 c_2 +d\, c_3 =0 \, , \quad c_1 + 4 c_2 + 2 c_3 =0 \,.
\eea
The above constraints fix the relative coefficients of the three tensors in the ansatz, and finally the Ward identity result (\ref{gT}) fixes completely the leftover overall normalization. 

 \begin{table}[t]
\centering 
\begin{tabular}{c|ccccccccccc}
{$T^{(m)}_{\mu \nu}$} $\backslash$  {constraints} &  { symmetric }  & { traceless } & {conservation} & { Ward identity } 
 \\ \hline
$T^{(2)}_{\mu \nu}$ &  $3$ & $2$ & $1$ & $0$ 
  \\
$T^{(4)}_{\mu \nu}$ &  $8$ & $5$ & $1$ & $0$ 
  \\
$T^{(6)}_{\mu \nu}$ &  $15$ & $10$ & $2$ & $0$ 
  \\
$T^{(8)}_{\mu \nu}$ &  $24$ & $16$ & $3$ & $0$
\end{tabular}
\caption{Here we show how energy momentum tensors with different numbers of derivatives are constructed by imposing various constraints. In each column, the numbers indicate how many free parameters of independent tensor structures are left after imposing the corresponding constraint shown on the top of that column. 
\label{tab:stresstensor}}
\end{table}

The same procedure can be applied to the non-trivial cases of energy momentum tensors  with higher derivatives. In particular, for the energy momentum tensor with four derivatives, we find eight independent symmetric tensors, the traceless condition reduces them to five, and the conservation fixes all the coefficients in front of each tensor except of an overall normalization. The latter is determined by the Ward Identity (\ref{gT}). As the result, we obtain, 
\begin{align}
 \label{eq:fourderivative}
T^{(4)}_{\mu \nu} (x) &= \frac{1}{2(d-1)} \biggr( (d{+}2)(\dd_\mu\sigma\dd_\nu\dd^2\sigma{+}\dd_\mu\dd^2\sigma\dd_\nu\sigma) {-} \frac{d(d{+}2)}{d{-}2}\dd_\mu\dd_\nu\sigma\dd^2\sigma{+}(d{-}4) \sigma \dd_\mu\dd_\nu\dd^2\sigma \cr
&{+}\frac{4d}{d{-}2}\dd_\mu\dd_\lambda\sigma\dd_\nu\dd^\lambda\sigma {-}4\,\dd_\mu\dd_\nu\dd_\lambda\sigma\dd^\lambda\sigma{+}\eta_{\mu\nu} \big(2\dd_\lambda\sigma\dd^\lambda\dd^2\sigma {+}\frac{d{+}2}{d{-}2}\dd^2\sigma\dd^2\sigma {-}\frac{4}{d{-}2} \dd_\lambda\dd_\rho\sigma \dd^\lambda\dd^\rho \sigma\big) \biggr)\,.
\end{align}
For $\s_2(x)$ this is relevant in $d=8$ and for $\s_1(x)$ in $d=6$.  

We have constructed the energy momentum tensors up to eight derivatives in this way. Some details of the construction may be found in table \ref{tab:stresstensor}, where we show the numbers of independent tensors at each step of imposing a particular constraint. We find, at least for the cases we have studied, the constraints are strong enough to completely fix all the free parameters. The results are presented in the Appendix~\ref{AppE}.

With the explicit forms of energy momentum tensors at hand, it is now straightforward to compute their two-point functions, and to determine $c_T$. To do so we use the following free scalar two-point function
\bea
\langle  \sigma(x_1) \sigma(x_2) \rangle  =  {N_{d} \over ( x^2_{12} )^{\Delta_{\sigma} }  } \, ,
\eea
with $\Delta_\sigma=\frac{d}{2}-n$, for a free theory with a $2n$-derivative energy momentum tensor $T^{(2n)}_{\mu \nu}$. The normalization factor $c_{d}$  is determined to be
\bea
N_{d}  = \frac{\Gamma\left(\frac{d}{2}-n\right)}{4^n\pi^{d/2}\Gamma(n)} \, ,
\eea
by requiring that the momentum space two-point function is normalized to unity, namely 
\bea
\langle  \sigma(p) \sigma(0) \rangle = \frac{1}{p^{\frac{d}{2}-\D}} \, ,
\eea
 $c^{(4)}_T$ and $c^{(6)}_T$ have also been recently computed in \cite{Osborn:2016bev}. As pointed out in \cite{Osborn:2016bev}, one may further simplify the task using the fact that the two-point function of energy momentum tensors is invariant by changing $T_{\rho \alpha} \rightarrow T_{\rho \alpha} + \partial^{\tau} X_{\tau \rho \alpha}$ for local $X_{\tau \rho \alpha}$, which leads to 
\bea
\langle  T^{(4)}_{\mu \nu}(x_1) T^{(4)}_{\rho \alpha}(x_2) \rangle =
-2 \langle  T^{(4)}_{\mu \nu}(x_1) \partial_{\rho} \partial_{\alpha} \sigma(x_2) \partial^2 \sigma (x_2) \rangle \, ,
\eea
and the right-hand side of the above equation is clearly easier to evaluate. As a consistency check we have computed the two-point function from both sides of the above equation using the energy momentum tensor (\ref{eq:fourderivative}), and we have obtain the same result, that reads 
\bea
c^{(4)}_T(d) = - {2 d (d+4) \over (d-1)(d-2)} \, .
\eea
Similarly for the six-derivative energy momentum tensor $T^{(6)}_{\mu \nu}$ in (\ref{eq:fourderivative}), we have the identity, 
\bea
\langle  T^{(6)}_{\mu \nu}(x_1) T^{(6)}_{\rho \alpha}(x_2) \rangle =
3 \langle  T^{(6)}_{\mu \nu}(x_1) \partial_{\rho} \partial_{\alpha} \sigma(x_2) \partial^4 \sigma (x_2) \rangle
\eea
and an explicit computation yields, 
\bea
c^{(6)}_T(d) = {3 d (d+4) (d+6) \over (d-4)(d-2)(d-1)} \, .
\eea
The results $c^{(4)}_T, c^{(6)}_T$ we obtained above using the energy momentum tensors fixed by constraints listed in table \ref{tab:stresstensor} are in agreement\footnote{Since we work on-shell our formulae for the energy momentum tensors differ from the corresponding ones of ~\cite{Osborn:2016bev}, however this is irrelevant to the final results for the $c_T(d)$'s.} with the ones in ~\cite{Diab:2016spb}. It is clear that the same procedure can be generalized and applicable to energy momentum tensors with higher derivatives. In particular, we have constructed the eight-derivative energy momentum tensor shown in (\ref{eq:eightderivative}) where we have fixed completely the coefficients of all the $24$ independent symmetric tensors using conservation, tracelessness and the Ward identity. Similarly the two-point function can be simplified as
\bea
\langle  T^{(8)}_{\mu \nu}(x_1) T^{(8)}_{\rho \alpha}(x_2) \rangle =
-4 \langle  T^{(8)}_{\mu \nu}(x_1) \partial_{\rho} \partial_{\alpha} \sigma(x_2) \partial^6 \sigma (x_2) \rangle \, . 
\eea
Our computation using $T^{(8)}_{\mu \nu}$ in (\ref{eq:eightderivative}) gives
\bea
c^{(8)}_T(d) = -\frac{4d(d{+}4)(d{+}6)(d{+}8)}{(d{-}1)(d{-}2)(d{-}4)(d{-}6)} \, .
\eea
This new result matches  the conjectured expression of the general $c^{(2k)}_T$ for a $2k$-derivative energy momentum tensor given in \cite{Osborn:2016bev}. When $d=12$ it yields $c_T$ and when $d=10$ it yields $\tilde{c}_T$ as
\begin{align}
c_T &=-{576 \over 11} \quad {\rm for}  \quad d=12 \, , \\
\tilde{c}_T &= -{280 \over 3} \quad {\rm for}  \quad d=10 \, .
\end{align}
These agree with the general formula given in \cite{Diab:2016spb}. 

\section{Outlook} \label{section:outlook}
In this preliminary work we have presented some basic features of the free $\s$CFTs. These are the theories describing the free dynamics of  $\s_2(x)$ and $\s_1(x)$ which are the shadows fields of the composites $\phi^2(x)$ and $\bar\psi(x)\psi(x)$ in free $d$ dimensional bosonic and fermionic CFTs. We have argued that  $\s_2(x)$ and $\s_1(x)$ can be defined as elementary free conformal fields for all even $d\geq 6$ and $d\geq 4$ respectively, and therefore $\s$CFTs can be studied using standard CFT methods. For $d>6$ and $d>4$ these theories are necessarily non-unitary and contain ghost and null fields in their spectrum. However, in the strict $d=\infty$ limit their spectrum coincides with that of the usual six- and four-dimensional free bosonic CFTs. As a nontrivial test of the above we have calculated the $c_T$ coefficients of both theories in $d=6,8,10$ using the conformal OPE, and found agreement with the more general results presented in \cite{Diab:2016spb}. For completeness, we have also calculated the $c_T$s constructing explicitly a symmetric, conserved and traceless energy momentum tensor in $d=6,8,10,12$. We again find agreement with \cite{Diab:2016spb} and \cite{Osborn:2016bev}.

We believe that a wealth of interesting directions opens up in the study of $\s$CFTs. We articulate a few of them below.
\begin{itemize}
\item
From (\ref{CTb}) and (\ref{CTf}) it is tempting to interpret the $c_T$'s, for $d$ even,  as some spin-dependent  shift in $N$. Indeed, for a given $d$ we can see from formulae like (\ref{d10s2calc}) and (\ref{d8s1calc}) that the $c_T$'s sum contributions from fields with spins up to $d-4$ for $\s_2$CFT and $d-2$ for the $\s_1$CFT. 
\item
An appealing property of the $\s$CFTs is the fact that $1/d$ naturally emerges as a small expansion parameter, actually always in the form $v/d$. This raises the interesting possibility of an $1/d$ expansion near the massless free CFTs in $d=6$ and $d=4$. In fact, it may be possible to organize a large-$d$ expansion near even-$d$ in all dimensions.
\item
It would also be interesting to go beyond the free $\s$CFTs, for example using the new bootstrap or the older skeleton expansion. This probably would involve moving away from $d$ even. 
\item
The scalar fields, like (\ref{O4}) that need to be included in the OPE in higher dimensions appear to be related to trace anomalies in the presence of scalar fields \cite{Osborn:1993cr,Bzowski:2015pba}. It would be nice to study this connection further.
\item
Finding a holographic dual for the $\s$CFTs would also be interesting. Scalar fields in AdS$_{d+1}$ with mass $m^2=2(2-d)$ and $m^2=1-d$ would give rise to pairs of boundary operators with dimensions $(d-2, 2)$ and $(d-1,1)$ respectively. They are above the Breitenlohner-Freedman bound for any $d$, and are both conformally coupled scalars in AdS$_4$, but only the first one in AdS$_6$. To describe holographically free $\s$CFTs one would need to consider even $d\geq 4$ and a higher-spin gauge theory. 

\end{itemize}

\section*{Acknowledgements}
A. C. P. wishes to thank the String Theory Group in the University of Rome, "Tor Vergata" for the warm hospitality extended to him during his visiting professorship and for the excellent scientific environment that initiated this work. The work of A. C. P. is partially supported by the MPNS–COST Action MP1210 “The String Theory Universe”. A. G. acknowledges the support by MPNS–COST Action MP1210 “The String Theory Universe” in the form of a STMS grant to visit Thessaloniki, where this work was finalized. We wish to thank M. Bianchi and A. Tseytlin for useful discussions and correspondence. 

\vspace{-0.3cm}

\appendix
\section{Scalar and energy momentum tensor correlation functions}
\label{AppA}
The two- and three-point functions of a scalar field $\O(x)$ and the energy momentum tensor $T_{\m\n}(x)$ are
\be
\label{OOTT}
\langle \O(x_1)\O(x_2)\rangle =C_{2}\frac{1}{x_{12}^{2\Delta_\O}}\,,\,\,\,\langle T_{\m\n}(x_1)T_{\r\s}(x_2)\rangle = C_{T}\frac{{\cal I}_{\m\n,\r\s}(x_{12})}{x_{12}^{2\Delta_{T}}} \, ,
\ee
where
\be
\label{I}
{\cal I}_{\m\n,\r\s}(x)=\frac{1}{2}\left[I_{\m\r}(x)I_{\n\s}(x)+I_{\m\s}(x)I_{\n\r}(x)\right]-\frac{1}{d}\delta_{\m\n}\delta_{\r\s}\,,\,\,\,I_{\m\n}(x)=\delta_{\m\n}-2\frac{x_\m x_\n}{x^2} \, .
\ee
The relevant three-point functions are given by 
\begin{align}
\label{phiphiO}
\langle \phi(x_1)\phi(x_2){\cal O}(x_3)\rangle &= \frac{g}{(x_{12}^2)^{\Delta_\phi-\frac{\D_\O}{2}}(x_{13}^2x_{23}^2)^{\frac{\D_\O}{2}}} \, , \\
\label{phiphiT}
\langle \phi(x_1)\phi(x_2)T_{\m\n}(x_3)\rangle &= \frac{g_{T}}{(x_{12}^2)^{\Delta_\phi-\frac{\D_{T}}{2}+1}(x_{13}^2x_{23}^2)^{\frac{\Delta_{T}}{2}+1}}\left[\left(X_{12}\right)_\m\left(X_{12}\right)_\n-\frac{1}{d}\eta_{\m\n}\left(X_{12}\right)^2\right] \, ,\nonumber \\
\left(X_{12}\right)_{\m}&=\frac{(x_{13})_\m}{x_{13}^2}-\frac{(x_{23})_\m}{x_{23}^2} \, .
\end{align}
The Ward identity for $T_{\m\n}$  fixes
\be
\label{gT}
g_T=\frac{d\Delta_\phi}{(d-1)S_d}\,.
\ee

\section{The CPWs for spin-$s$ fields}
\label{AppB}
The irreps of $SO(d,2)$ with spin-$s$ and dimension $\D_s$ correspond to totally symmetric and traceless rank-$s$ tensors $T_{\m_1\m_2 \ldots \m_s}$. Their contribution to the four-point function of scalar fields has been known for a long time \cite{Hoffmann:2000mx}. Let us recall it briefly here. The three-point function of a scalar field ${\cal O}$ with dimension $\D_{\O}$ with these irreducible tensors are given by
\begin{align}
\label{OOT}
\langle\O(x_1)\O(_2)T_{\m_1\m_2 \ldots \m_s}(x)\rangle &= g_{T_s}\frac{{\cal M}_{\m_1\m_2 \ldots \m_s}(x_1,x_2;x_0)}{(x_{12}^2)^{\D_\O-\frac{\D_{s}}{2}}(x_{10}^2 x_{20}^2)^{\frac{\D_s}{2}}} \, ,\\
\label{M}
{\cal M}_{\m_1\m_2 \ldots \m_s} (x_1,x_2;x_0)&=\left[e_{\m_1}e_{\m_2} \ldots e_{\m_s}-{ \rm traces}\right] \, .
\end{align}
We have denoted
\be
\label{es}
e_{\m}=\frac{\xi_\m}{|\xi^2|^{1/2}}\,,\,\,\,\xi_{\m}=\frac{(x_{10})_\m}{x_{10}^2}-\frac{(x_{20})_\m}{x_{20}^2}\,,\,\,\,\xi^2=\xi\cdot\xi=\frac{x_{12}^2}{x_{10}^2x_{20}^2}\,,\,\, e\cdot e=1
\ee
Let us give two examples. For spin-2 the tensor structure on the RHS of (\ref{OOT}) is
\be
\label{3ptspin2}
{\cal M}_{\m\n}(x_1,x_2;x_0)=e_\m e_\n-\frac{1}{d}\delta_{\m\n} \, .
\ee
For spin-4 we find
\begin{align}
\label{3prspin4}
{\cal M}_{\m\n\rho\sigma}(x_1,x_2;x_0) =& \,\,e_\m e_\n e_\rho e_\sigma-\frac{1}{d+4}\left(\delta_{\m\n}e_\rho e_\sigma+\delta_{\m\rho} e_{\n}e_{\sigma}+\delta_{\m\sigma} e_{\rho} e_{\nu}+\delta_{\n\rho} e_{\m} e_{\sigma}+\delta_{\n\sigma} e_{\m} e_{\rho}+\delta_{\rho\sigma} e_{\m} e_{\n}\right)\nonumber\\
&+\frac{1}{(d+2)(d+4)}\left(\delta_{\m\n}\delta_{\rho\sigma}+\delta_{\m\rho}\delta_{\n\sigma}+\delta_{\m\sigma}\delta_{\n\rho}\right)
\, .
\end{align}
The CPW of $T_{\m_1\m_2 \ldots \m_s}$ is proportional to the {\it canonical}\footnote{This is found by dropping the {\it shadow} contribution of the integral. It should be noted however that the shadow contribution is necessary to ensure crossing symmetry when  exchange integrals like (\ref{beta}) turn up in field theoretic CFT calculations \cite{Hoffmann:2000tr}.} part of the following exchange integral
\begin{align}
\label{beta}
\b_{\D_\O,\D_s}(x_1,x_2,x_3,x_4)&=\int d^d x_0\langle\O(x_1)\O(x_2)T_{\m_1 \ldots \m_s}(x_0)\rangle\langle\tilde{T}_{\m_1 \ldots \m_s}(x_0)\O(x_3)\O(x_4)\rangle\nonumber\\
&\hspace{-2cm}=\frac{g_{T_s}g_{\tilde{T}_s}}{(x_{12}^2x_{34}^2)^{\D_\O}}\frac{(x_{12}^2)^{\frac{\D_{s}}{2}}}{(x_{34}^2)^{\frac{\D_{s}}{2}-\frac{d}{2}}}\int d^dx\frac{{\cal M}_{\m_1 \ldots \m_s}(x_1,x_2;x_0){\cal M}_{\m_1 \ldots \m_s}(x_3,x_4;x_0)}{(x_{10}^2x_{20}^2)^{\frac{\D_s}{2}}(x_{30}^2x_{40}^2)^{\frac{d}{2}-\frac{\D_s}{2}}} \, ,
\end{align}
where $\tilde{T}_{\m_1\m_2 \ldots \m_s}$ is the {\it shadow} spin-$s$ irrep represented by a symmetric traceless rank-$s$ tensor field with dimension $\tilde{\D}_s=d-\D_{s}$. To perform the integral in the last line of (\ref{beta}) we use the fact, proven for example in \cite{Hoffmann:2000mx}, that
\be
{\cal M}_{\m_1, \ldots , \m_{s}}(x_1,x_2;x_0){\cal M}_{\m_1, \ldots ,\m_s}(x_3,x_4;x_0)=\frac{1}{c^{(s)}_{s;d}}C_{s}^{\frac{d}{2}-1}(t)\,,\,\,\,t=e\cdot e' \, ,
\ee
where $e'_\m$ is gotten by substituting $x_1\mapsto x_3$ and $x_2\mapsto x_4$ in (\ref{es}) and $C_{s}^{\frac{d}{2}-1}(t)$ is the Gegenbauer polynomial in the variable $t$ defined as
\be
\label{Gegen}
C_{s}^{\frac{d}{2}-1}(t)=\sum_{n=0}^{s}c_{s;d}^{(n)}t^n \, .
\ee
The relevant examples for $s=2,4$ and $6$ are
\begin{align}
\label{Gegen24}
\frac{1}{c_{2;d}^{(2)}}C_2^{\frac{d}{2}-1}(t)&=t^2-\frac{1}{d}\,,\,\,\,\frac{1}{c_{4;d}^{(4)}}C_4^{\frac{d}{2}-1}(t)=t^4-\frac{6}{d+4}t^2+\frac{3}{(d+2)(d+4)} \, ,\\
\label{Gegen6}
\frac{1}{c_{6;d}^{(6)}}C_6^{\frac{d}{2}-1}(t) &=t^6-\frac{15}{d+8}t^4+\frac{45}{(d+8)(d+6)}t^2-\frac{15}{(d+4)(d+6)(d+8)} \, .
\end{align}
For a given $s$ the leading-$d$ result is given by the $t^s$ term in the expansion (\ref{Gegen}). Let us give the $s=2$ example, denoting the dimension of the corresponding field $\D_{2}$. 
We obtain
\bea
\label{MtimesM}
{\cal M}_{\mu \nu}(x_1,x_2;x_0){\cal M}_{\m\n}(x_3,x_4;x_0)  &=& 
\left[ 
\left( {x_{10} \over x^2_{10} } - {x_{20} \over x^2_{20}}  \right) \cdot \left( {x_{30} \over x^2_{30} } - {x_{40} \over x^2_{40}}  \right)
\right]^2 - {1 \over d} {x^2_{12} x^2_{34} \over x^2_{10}x^2_{20}x^2_{30}x^2_{40} } \cr
&=& 
{1 \over 4}{ N^2 \over \left(x^2_{10} x^2_{20} x^2_{30} x^2_{40}\right)^2 }  - {1 \over d} {x^2_{12} x^2_{34} \over x^2_{10}x^2_{20}x^2_{30}x^2_{40} } \, ,
\eea
and the numerator $N$ is given by
\bea
{N^2 \over \left(x^2_{10} x^2_{20} x^2_{30} x^2_{40}\right)^2 }
&=&\,\,
{1 \over 4} \left[ {x_{13}^4 \over x_{10}^4 x_{30}^4 } 
+ {x_{23}^4 \over  x_{20}^4 x_{30}^4} + 
{x_{14}^4 \over x_{10}^4 x_{40}^4 } + 
{ x_{24}^4 \over x_{20}^4 x_{40}^4}  +
{2 x_{13}^2 x_{24}^2 \over x_{10}^2 x_{20}^2 x_{30}^2 x_{40}^2 } + 
  { 2 x_{14}^2 x_{23}^2 \over x_{10}^2 x_{20}^2 x_{30}^2 x_{40}^2} \right.
 \cr
&-& \left.
{ 2 x_{13}^2 x_{23}^2 \over x_{10}^2 x_{20}^2 x_{30}^4 } -  
{2x_{14}^2 x_{24}^2 \over  x_{10}^2 x_{20}^2 x_{40}^4 } 
 -  {2 x_{13}^2 x_{14}^2 \over x_{10}^4 x_{30}^2 x_{40}^2 }
  - 
  {2 x_{24}^2 x_{23}^2 \over x_{20}^4 x_{30}^2 x_{40}^2 } \right] \, .
\eea
We then notice that all the integrals involved in (\ref{beta}) are of the conformal 4-star form\footnote{We  have dropped the $x$-dependance to simplify notation.}
\begin{align}
\int d^d x_0 {1 \over x^{2 a_1}_{10} x^{2 a_2}_{20} x^{2 a_3}_{30} x^{2 a_4}_{40} } = S(a_1,a_2,a_3,a_4)+S_{shadow}(a_1,a_2,a_3,a_4) \, ,
\end{align}
with a constraint $a_1 + a_2 +a_3 + a_4 = d$ and the two conditions $a_1+a_2=\D_2$, $a_3+a_4=d-\D_2$. This integral has been calculated many times as a double series in $v$ and $Y$. Its canonical part reads as
\be
S(a_1,a_2,a_3,a_4)
=
\pi^{\frac{d}{2}}\frac{\Gamma\left(\frac{d}{2} -\D_2\right) }{\Gamma(\D_2)}{{\cal S}_{a_1,a_2,a_3,a_4}\over (x_{23}^2)^{\frac{d}{2} - a_4} (x^{2}_{14})^{a_1} (x_{34}^{2})^{\frac{d}{2}-\D_2}
(x_{24}^{2})^{a_2 + a_4 -\frac{d}{2}}} \, ,
\ee
with
\begin{align}
{\cal S}_{a_1,a_2,a_3,a_4}&=\sum^{\infty}_{n=0} {v^n \over n!} \frac{A^{(n)}_{a_1,a_2,a_3,a_4}}{\left(\D_2\right)_{2n}\left(\D_2+1-\frac{d}{2} \right)_n }{}_2F_1\left(\frac{d}{2}-a_4+n, a_1 +n;\D_2+2n;Y\right)\,,
\\
A^{(n)}_{a_1,a_2,a_3,a_4} &={\Gamma\left(\frac{d}{2}-a_3+n\right) \Gamma\left(a_2+n\right)\Gamma\left(\frac{d}{2} -a_4+n\right) 
\Gamma\left(a_1 +n\right)  \over \left[\prod^4_{i=1}\Gamma(a_i)\right]  } \, .
\end{align}
Notice at first that the $O(1/d)$ term in (\ref{MtimesM}) gives by virtue of (\ref{beta}) the CPW of a scalar with dimension $\D_{2}$. This is the {\it seed} CPW and we can denote the corresponding ${\cal S}$-function as
\be
\label{seedSfunction}
{\cal S}_{\frac{\D_{2}}{2},\frac{\D_{2}}{2},\frac{d}{2}-\frac{\D_{2}}{2},\frac{d}{2}-\frac{\D_{2}}{2}} \equiv {\cal S}_{\D_{2}} \, .
\ee
All other ${\cal S}$-functions that are involved in the CPW are obtained from (\ref{seedSfunction}) by raising/lowering by 1 the indices $a_1, \ldots , a_4$. We can define an operator $\hat{L}$  that does the job
\be
\label{Loper}
\hat{L}_{k_1,k_2,k_3,k_4}\cdot {\cal S}_{a_1,a_2,a_3,a_4}={\cal S}_{a_1+k_1,a_2+k_2,a_3+k_3,a_4+k_4} \, ,
\ee
 for the integers $k_1, \ldots ,k_4$. So, with the above machinery the CPW of the energy momentum tensor can be expressed as
\bea
\label{Hs2}
\cH_{\Delta_{2}}(v,Y)
&=&\,\, {\cal C}_2 v^{\frac{\D_2}{2}-1}
\left[\frac{1}{4}\left\{ \hL_{1,-1,1,-1}+\hL_{1,-1,-1,1}+\hL_{-1,1,1,-1}+\hL_{-1,1,-1,1}\right\}\right.
\cr
&-& 
{1 \over 2}\left\{\hL_{1,-1,0,0}+\hL_{-1,1,0,0}+\hL_{0,0,1,-1}+\hL_{0,0,-1,1}+2\mathbb{I}\right\}\cr
&-&\left.\frac{1}{2}Y\left\{\hL_{1,-1,1,-1}-\hL_{1,-1,0,0}-\hL_{0,0,1,-1}+\mathbb{I}\right\}+\frac{1}{4}Y^2\hL_{1,-1,1,-1}-\frac{1}{d}v\mathbb{I}\right]\cdot{\cal S}_{\D_2}\cr
&=&
\,\, {\cal C}_2 v^{\frac{\D_2}{2}-1}
\left[\hat{t}^2-\frac{1}{d}v\mathbb{I}\right]\cdot{\cal S}_{\D_2} \, ,
\eea
with $\mathbb{I}$ the unit operator. The overall constant ${\cal C}_2$ is fixed by comparing with the explicit four-point function result. Formula (\ref{Hs2}) serves also as the definition of the operator $\hat{t}^2$. Comparing with (\ref{Gegen24}), (\ref{Gegen6}) we see that the general CPW of the spin-$s$ field with dimension $\D_s$ can be formally written as
\bea
\label{Hsgeneral}
H_{\D_s}(v,Y)={\cal C}_s v^{\frac{\D_s}{2}-\frac{s}{2}}
\frac{1}{c_s^{(s;d)}}C_{s}^{\frac{d}{2}-1}(\hat{t})\cdot{\cal S}_{\D_s} \, ,
\eea
where the higher powers of $\hat{t}$ are obtained using 
\be
\hat{L}_{k_1,k_2,k_3,k_4}\cdot\hat{L}_{m_1,m_2,m_3,m_4}=\hat{L}_{k_1+m_1,k_2+m_2,k_3+m_3,k_4+m_4}\, ,
\ee
As we consider even spin $s$, we will be interested in the action of $\hat{t}^2$ on ${\cal S}_\D$. For that we note the following properties
\be
\label{t2S}
(\hat t^{2} +Y{-}2)F_\D^0\left(Y\right) =2(Y{-}1)\,_2F_1\left(\frac{\Delta}{2}{+}1,\frac{\Delta}{2};\Delta;Y\right)\, ,
\ee
Using then the following two identities 
\bea
_2F_1\left(\frac{\Delta}{2}{+}1,\frac{\Delta}{2};\Delta;Y\right)
&{=}&
\frac{1}{\Delta(\Delta{+}1)Y^{\Delta-1}}\frac{d}{dY}\left[ (1{-}Y)^{\frac{\Delta}{2}{-}1}\frac{d}{dY}\left(\frac{Y^{\Delta{+}1}}{(1{-}Y)^{\frac{\Delta}{2}}}\,F_\D^1\left(Y\right)\right) \right]\,,\\
F_\D^0\left(Y\right)
&{=}&
\frac{1}{\Delta(\Delta{+}1)Y^{\Delta-1}}\frac{d}{dY}\left[ (1{-}Y)\frac{d}{dY}\left(Y^{\Delta{+}1}\,F_\D^1\left(Y\right)\right) \right],
\eea
we get from (\ref{t2S}) (using now the compact notation introduced in the text)  
\be
\hat t^{2}\cdot F_\D^0(Y)=\frac{Y^2}{4}\frac{\Delta}{\Delta+1}F_{\D+2}^0(Y)\equiv \frac{Y^2}{4}\frac{\Delta}{\Delta+1}F_{\D}^1(Y) \, .
\label{t2rule}
\ee
In other words, the action of the operator $\hat t^{2}$ on a hypergeometric function of the form ${}_2F_1(a,a,2a,Y)$ is to shift the parameter $a\to a{+}1$ and to multiply by an overall constant times $Y^2$. The leading $d$ behavior of (\ref{Hsgeneral}) is determined by the action of the operator $(\hat t^{2})^{s/2}$, therefore, using (\ref{t2rule}), we prove that the $v$-expansion of the spin-$s$ CPW starts with
\be
(\hat  t^{2})^{s/2}\cdot F_{\D_s}^0(Y)\sim Y^s F_{\D_s+s}^0(Y)=Y^s F_{\D_s}^{s/2}(Y) \, . \ee
\section{Expansion of the spin-2 conformal partial wave}
\label{AppC}
In this section we consider the full $v$ expansion of the spin-2 CPW. The coefficient of order $(v/d)^n$ in the expansion of  ${\cal S}_{\Delta_2}$ is
\be
{\cal S}^{(n)}_{\Delta_2}(Y){=}\frac{v^n}{n!}\frac{\Gamma\left(\frac{\Delta_2}{2}{+}n\right)^4 }{\Gamma\left(\frac{\Delta_2}{2}\right)^2\Gamma\left(\frac{d}{2}{-}\frac{\Delta_2}{2}\right)^2} \frac{ 1}{\left(1{-}\frac{d}{2}{+}\Delta_2\right)_n\left(\Delta_2\right)_{2n}}F_{\D_2}^n(Y) \, .
\ee
Therefore, the coefficient of order $(v/d)^n$ in the expansion of the spin-2 CPW is given by
\be
\left(\frac{v}{d}\right)^n H_{\D_2}^{(n)}(Y,d;s)=\hat t^{(2)}\cdot{\cal S}^{(n)}_{\Delta_2}(Y)-\frac{v}{d}{\cal S}^{(n{-}1)}_{\Delta_2}(Y)\, .
\ee
Expanding this coefficient order by order in $Y$ we recognize that the spin-2 CPW can be written as follows
\be
\mathcal{H}_{\Delta_2}^{(2)}(v,Y) = v^{\frac{\Delta_2-2}{2}}\sum_{n=0}^{\infty} v^n \,\k_{\D_2,2n}\left[Y^2F_{\D_2+2}^n(Y) -4\frac{v}{d}\alpha_{\D_2,2n}^{(1)}F_{\D_2}^n(Y)\right] \, ,
\ee
where
\begin{align}
\k_{\D_2,2n}&=\frac{\Delta_2(\Delta_2{-}1)}{2^{2(n{+}2)}n!}\frac{\left( \frac{\Delta_2}{2} {+}1\right)_n^2\left( \frac{\Delta_2}{2}{+}1 \right)_{n{-}1}}{\left( \Delta_2{+}1{-}\frac{d}{2} \right)_n\left( \frac{\Delta_2{+}1}{2}{+}1 \right)_{n{-}1}\left( \frac{\Delta_2{+}1}{2}{+}n{-}1 \right)_2} \, ,\label{k2}\\
\alpha_{\D_2,2n}^{(1)}&=\frac{(\Delta_2{-}1{+}2n)(\Delta^2(\Delta{+}1{+}2n) -\Delta_2(\Delta_2{+}2n)\,d +2n\,d)}{(\Delta_2{-}1)(\Delta_2{+}1{-}d)(\Delta_2{+}2n)^2} \, .
\label{alpha2}
\end{align}

\section{Expansion of the spin-4 conformal partial wave}
\label{AppD}
Similarly with  the spin-2 CPW, the coefficient of order $(v/d)^n$ of the spin-4 CPW is given by
\be
\left(\frac{v}{d}\right)^n H_{\D_4}^{(n)}(Y,d;s)=\frac{1}{{\cal C}(d,\Delta_4)}\left( (\hat t^{2})^2\cdot{\cal S}^{(n)}_{\Delta_4}-\frac{6\,v}{d{+}4} \hat{t}^2\cdot{\cal S}^{(n{-}1)}_{\Delta_4}+\frac{3\,v^2}{(d{+}2)(d{+}4)}{\cal S}^{(n{-}2)}_{\Delta_4}\right).
\ee
Therefore the spin-4 CPW can be written as
\begin{align}
\mathcal{H}_{\Delta_4}^{(4)}(v,Y)=v^{\frac{\Delta_4-4}{2}}\sum_{n=0}^{\infty} v^n \,\k_{\D_4,2n} \biggr[&Y^4F_{\D_4+4}^n(Y)-4\frac{6\,v}{d{+}4}\frac{Y^2}{4}\alpha^{(1)}_{\D_4,2n}F_{\D_4+2}^n(Y)+4^2 \frac{3\,v^2}{(d{+}2)(d{+}4)}\alpha^{(2)}_{\D_4,2n}F_{\D_4}^n(Y)\biggr] \, ,
\end{align}
with
\begin{align}
\k_{\D_4,2n}&=\frac{(\Delta_4{-}1)\Delta_4(\Delta_4{+}1)}{2^{2n{+}5}n!}\frac{\left(\frac{\Delta_4}{2}{+}2\right)_n^2\left(\frac{\Delta_4}{2}{+}1\right)_{n{-}1}}{\left(\Delta_4{+}1{-}\frac{d}{2}\right)_n\left(\frac{\Delta_4{+}1}{2}{+}2\right)_{n{-}1}\left(\frac{\Delta_4{+}1}{2}{+}n{-}1\right)_3}\,,\\
\alpha^{(1)}_{\D_4,2n}&=\frac{(\Delta_4{+}1{+}2n)(\Delta_4^2(\Delta_4{+}3{+}2n)-\Delta_4(\Delta_4{+}4{+}2n)\,d-(2n{+}4)\,d-4)}{(\Delta_4{+}1)(\Delta_4{-}1{-}d)(\Delta_4{+}2{+}2n)^2} \, , \\
\alpha^{(2)}_{\D_4,2n}&=\frac{((\Delta_4{+}2n)^2{-}1)\,\mathcal{P}_n(\Delta_4,d)}{((\Delta_4{-}d)^2{-}1)(\Delta_4{-}1)(\Delta_4{+}1)(\Delta_4{+}2n)^2(\Delta_4{+}2{+}2n)^2} \, ,
\end{align}
where 
\bea
\mathcal{P}_n(\Delta_4,d)&{=}&\Delta_4^2(\Delta_4{+}2)^2(\Delta_4^2{-}1)+4n^2(\Delta_4^4{+}8)+4n\Delta_4(\Delta_4^4{+}2\Delta_4^3{-}2\Delta_4{+}8)\cr
&{-}&2d(\Delta_4^3(\Delta_4{+}2)^2+4n^2(\Delta_4^3{-}1)+2n(6{+}\Delta_4({-}4{+}\Delta_4(1{+}2\Delta_4(\Delta_4{+}2)))))\cr
&{+}&d^2(\Delta_4^4+4(n{+}1)\Delta_4^3+4(n{+}1)^2\Delta_4^2-4n(n{+}3))\, .
\eea

\section{Six and eight-derivative energy momentum tensors}
\label{AppE}
The energy momentum tensor with six derivatives is given by
\bea  \label{eq:sixderivative}
T^{(6)}_{\mu \nu} (x)&=& \frac{1}{2(d{-}1)}\biggr(\!\!-(d{+}4)(\dd_\mu\sigma\dd_\nu\dd^4\sigma{+} ( \mu \leftrightarrow \nu) ){-}\frac{d(d{+}4)}{(d{-}4)}\dd_\mu\dd^2\sigma\dd_\nu\dd^2\sigma
{-}\frac{32d}{(d{-}2)(d{-}4)}\dd_\mu\dd_\lambda\dd_\rho\sigma\dd_\nu\dd^\lambda\dd^\rho\sigma \cr
&{-}&
\frac{8(d{+}2)}{d{-}4}(\dd_\mu\dd_\lambda\dd^2\sigma\dd_\nu\dd^\lambda\sigma{+} ( \mu \leftrightarrow \nu) ) {+}\frac{(d{-}2)(d{+}4)}{d{-}4}\dd_\mu\dd_\nu\dd^2\sigma\dd^2\sigma  
{+}
\frac{32}{d{-}4}\dd_\mu\dd_\nu\dd_\lambda\dd_\rho\sigma\dd^\lambda\dd^\rho\sigma \cr
&{+}&
\frac{(d{+}2)(d{+}4)}{d{-}4}\dd_\mu\dd_\nu\sigma\dd^4\sigma {+}\frac{8d(d{+}2)}{(d{-}2)(d{-}4)} \dd_\mu\dd_\nu\dd_\lambda\sigma\dd^\lambda\dd^2\sigma  {+}8\dd_\mu\dd_\nu\dd_\lambda\dd^2\sigma\dd^\lambda\sigma
{+}(d{-}6) \sigma \dd_\mu\dd_\nu\dd^4\sigma
 \cr
&{+}&
\eta_{\mu\nu}\big( 2\dd_\lambda\sigma\dd^\lambda\dd^4\sigma +\frac{d^2{-}6d{-}24}{(d{-}2)(d{-}4)}\dd_\lambda\dd^2\sigma\dd^\lambda\dd^2\sigma{+}\frac{16}{d{-}4}\dd_\lambda\dd_\rho\sigma\dd^\lambda\dd^\rho\dd^2\sigma
{-}\frac{2(d{+}4)}{d{-}4}\dd^2\sigma\dd^4\sigma \cr
&{+}&
\frac{32}{(d{-}2)(d{-}4)}\dd_\lambda\dd_\rho\dd_\alpha \sigma\dd^\lambda\dd^\rho\dd^\alpha \sigma \big)\biggr) \, ,
\eea
This is relevant for $\s_2(x)$ when $d=10$ and for $\s_1(x)$ when $d=8$.

The energy momentum tensor with eight derivatives takes the following form, 
\bea  \label{eq:eightderivative}
T^{(8)}_{\mu \nu} (x)
&=&
 \frac{1}{2(d{-}1)}\biggr( \frac{(d{+}2)(d{+}6)}{d{-}6}(\dd_\mu\dd^2\sigma \dd_\nu\dd^4\sigma{+} (\mu \leftrightarrow \nu) ) {+}(d{+}6)(\dd_\mu \sigma \dd_\nu\dd^6\sigma{+}(\mu \leftrightarrow \nu) )\cr
&{+}&\frac{12(d{+}4)}{(d{-}6)}
(\dd_\mu\dd_\lambda\sigma\dd_\nu\dd^\lambda\dd^4\sigma  {+}  ( \mu \leftrightarrow  \nu ) ){+}\frac{16d(d^2{+}2d{-}12)}{(d{-}2)(d{-}4)(d{-}6)}\dd_\mu\dd_\lambda\dd^2\sigma\dd_\nu\dd^\lambda\dd^2\sigma \cr
&{+}&
\frac{96(d{+}2)}{(d{-}4)(d{-}6)} (\dd_\mu\dd_\lambda\dd_\rho\dd^2\sigma \dd_\nu\dd^\lambda\dd^\rho\sigma{+} ( \mu \leftrightarrow  \nu ) ){+}
\frac{384d}{(d{-}2)(d{-}4)(d{-}6)}\dd_\mu\dd_\lambda\dd_\rho\dd_\alpha \sigma\dd_\nu\dd^\lambda\dd^\rho\dd^\alpha\sigma\cr
&{-}&
\frac{(d{-}4)(d{+}6)}{d{-}6}\dd_\mu\dd_\nu\dd^4\sigma\dd^2\sigma {-}\frac{d(d{+}2)(d{+}6)}{(d{-}2)(d{-}6)}\dd_\mu\dd_\nu\dd_\lambda\dd^2\sigma\dd^\lambda\dd^2\sigma {-}\frac{16(d^2{+}2d{-}12)}{(d{-}6)(d{-}4)}\dd_\mu\dd_\nu\dd^2\sigma\dd^4\sigma\cr
&{-}&
\frac{(d{+}4)(d{+}6)}{d{-}6}\dd_\mu\dd_\nu\sigma\dd^6\sigma {-}\frac{12(d{+}2)(d{+}4)}{(d{-}6)(d{-}4)} \dd_\mu\dd_\nu\dd_\lambda\sigma \dd^\lambda\dd^4\sigma {-}\frac{96d(d{+}2)}{(d{-}2)(d{-}4)(d{-}6)}\dd_\mu\dd_\nu\dd_\lambda\dd_\rho\sigma\dd^\lambda\dd^\rho\dd^2\sigma \cr
&{-}&
(d{-}8)\dd_\mu\dd_\nu\dd^6\sigma\,\sigma {-}12\,\dd_\mu\dd_\nu\dd_\lambda\dd^4\sigma\dd^\lambda\sigma {-}\frac{96}{d{-}6}\dd_\mu\dd_\nu\dd_\lambda\dd_\rho\dd^2\sigma\dd^\lambda\dd^\rho\sigma{-}\frac{384}{(d{-}4)(d{-}6)}\dd_\mu\dd_\nu\dd_\lambda\dd_\rho\dd_\alpha \sigma\dd^\lambda\dd^\rho\dd^\alpha \sigma\cr
&{+}&
\eta_{\mu\nu}\big( {-}\frac{2(d^2{-}10d{-}72)}{(d{-}4)(d{-}6)}\dd_\lambda\dd^2\sigma\dd^\lambda\dd^4\sigma {+}2\,\dd_\lambda\sigma\dd^\lambda\dd^6\sigma {-}\frac{24}{d{-}6}\dd_\lambda\dd_\rho\sigma\dd^\lambda\dd^\rho\dd^4\sigma 
{+}\frac{(d{+}2)(d{+}6)}{(d{-}2)(d{-}6)}\dd^4\sigma\dd^4\sigma
 \cr
&{-}&
\frac{192}{(d{-}4)(d{-}6)}\dd_\lambda\dd_\rho\dd_\alpha \sigma\dd^\lambda\dd^\rho\dd^\alpha \dd^2\sigma {-} \frac{384}{(d{-}2)(d{-}4)(d{-}6)}\dd_\lambda\dd_\rho\dd_\alpha \dd_\kappa\sigma\dd^\lambda\dd^\rho\dd^\alpha \dd^\kappa\sigma 
 \cr
&{+}&
\frac{2(d{+}6)}{d{-}6}\dd^2\sigma\dd^6\sigma
{-}\frac{16(d^2{-}4d{-}24)}{(d{-}2)(d{-}4)(d{-}6)}\dd_\lambda\dd_\rho\dd^2\sigma\dd^\lambda\dd^\rho\dd^2\sigma
\big)\biggr)  \, .
\eea
This is relevant for $\s_2(x)$ when $d=12$ and for $\s_1(x)$ when $d=10$.

\section{$c_T$ for genral $d$ from the renormalization of composites}
We present here an apparently ad hoc, but nevertheless intriguing argument that yields both the $c_T(d)$'s for any $d$.  The relationship between the elementary free field $\phi_d(x)$ in $d$ dimensions and $\sigma_2(x) $ can be calibrated in $d=6$ as
\be
\label{phisigma}
\Lambda^{6-d}\phi_d(x)\phi_d(x)=\sigma^2_2(x) \, .
\ee
For $d>6$ we take $\Lambda$ to be an IR cutoff such that for $\Lambda\rightarrow 0$ we look for an UV scaling limit. To see how the above operator relationship "runs" with $d$ we notice that since the scaling dimension of $\s_2(x)$ is fixed, the ratio $t=\frac{\Lambda^2}{\sigma}$ is a dimensionless RG parameter along $d$.  Hence we could define a renormalized field $\phi_{R,d}(x)$ as 
\be
\phi_d(x)=Z_\phi(t)\phi_{R,d}(x)=\left(1+\alpha_1t+\alpha_2t^2+ \ldots \right)\phi_{R,d}(x) \, .
\ee
Then (\ref{phisigma}) becomes
\be
\label{phiR}
\Lambda^{6-d}\left(1+\alpha_1t+\alpha_2t^2+ \ldots \right)^2\phi_R(d)\phi_R(d)=\sigma^2 \, .
\ee
For even $d$, picking the $\Lambda$-independent term on the LHS of (\ref{phiR}) we will find
\be
\label{phiRsigma}
c_d\phi_{R,d}(x)\phi_{R,d}(x)=\sigma^{\frac{d}{2}-1}(x)+ \cdots \, ,
\ee
where the dots denote $\Lambda$-dependent terms. A finite number of them are divergent as $\Lambda\rightarrow 0$ and the rest vanish in this limit. Curiously, if we choose
\be
Z(t)=\sum_{n=0}^{\infty}(-1)^nC(n)t^n \, , 
\ee
with $C(n)$ the Catalan numbers
\be
C(n)=\frac{(2n)!}{n!(n+1)!} \, ,
\ee
the $\Lambda$-independent term on the LHS of (\ref{phiR}) arises for $n=d/2-3$ and $d>6$. We obtain
\be
c_d=\frac{c_T(d)}{C_T^{(0)}(d)}=(-1)^{\frac{d}{2}-3}\left[C\left(\frac{d}{2}\right)-2C\left(\frac{d}{2}-1\right)\right] \, .
\ee
agreeing with the results in \cite{Diab:2016spb}.

The fermionic case works out the same way, but with a twist. Here we calibrate $\s_1(x)$ with the composite fermionic field $\psi_d(x)\psi_d(x)$ in $d=2$ as 
\be
\label{psi} 
\Lambda^{2-d}\frac{1}{4}\bar{\psi}_d(x)\psi_d(x)=\s_1(x) \, .
\ee
Now the RG parameter is $t=\Lambda/\sigma$ and renormalizing as above we have
\be \label{eq:RG2}
\psi_d(x)=Z_\phi(t)\psi_{R,d}(x)=(1+\alpha_2t^2+\alpha_4t^4+\alpha_6t^6+\cdots)\psi_{R,d}(x) \, ,
\ee
leading to $\Lambda$-independent term as
\be
\tilde{c}_{d}\bar\psi_{R,d}(x)\psi_{R,d}(x)=\s_1^{d-1}(x)+\cdots \, .
\ee
Remarkably, we can tune the $\alpha$'s to reproduce 
\be
\tilde{c}_d=-\frac{\tilde{c}_T(d)}{C_T^{(0)}(d)} \, ,
\ee
to any order we wish. For example, setting
\be \label{eq:alphainteger}
\alpha_2=-2\,,\alpha_4=8\,,\alpha_6=-26\,,\alpha_8=84\,,\alpha_{10}=-284\,,\alpha_{12}=996\,,\alpha_{14}=-3562\,,
\ldots \, ,
\ee
we obtain
\be
\label{tcd}
\tilde{c}_2=-\frac{1}{4}\,,\tilde{c}_4=1\,,\tilde{c}_6=-5\,,\tilde{c}_8=21\,,\tilde{c}_{10}=-84\,,\tilde{c}_{12}=330\,,\tilde{c}_{14}=-1287\,, \ldots \, . 
\ee
The numbers (\ref{tcd}) are, for even $d>4$, the integers in (4.28) of \cite{Diab:2016spb}. However, we are not aware of a simple combinatorial interpretation of the numbers in (\ref{eq:alphainteger}).

\bibliographystyle{JHEP}
\bibliography{Refs}

\end{document}